\documentclass[11pt,dvips]{article}
\usepackage{epsfig,times}
\usepackage[english]{babel}

\begin{document}
\begin{center}
{\bf \Large 
Initial fluctuations and correlation of finite distributions of secondary particles in interaction of heavy ions with photoemulsion nuclei 
} 
\vspace{2mm}

{\it \large A.Sh. Gaitinov, I.A. Lebedev, A.I. Fedosimova}

\vspace{2mm}

{\large Institute of Physics and Technology, Almaty, Kazakhstan}

\vspace{2mm}

\end{center}

{\small 
The study of the peculiarities of the distribution of secondary particles depending on the degree centrality and the degree of asymmetry of the interacting nuclei, is performed. The number of multicharged fragments of the projectile nucleus $N_f$  in interactions sharply asymmetric nuclei depends on the centrality degree of interaction. At that, the events with $N_f$ =1 are  separated clearly in the distribution of the total charge of the fragments of a projectile nucleus $Q$ depending on the nature of the correlation of the number of fragments of the target nucleus and the multiplicity of secondary particles.
}

\large

\section{Introduction}

The dynamics of the interaction of nuclei is determined not only by the energy and mass of the interacting nuclei, but also the geometry of the colliding nuclei. Accounting of the influence of fluctuations of the initial states in individual interactions allows to explore the true dynamic correlation of the final state of interactions of excited hadronic systems \cite{1}-\cite{3}.

In central collisions the number of generated secondary particles is maximal. If the collision is peripheral, then the overlap of the interacting nuclei is incomplete and resulting fireball expands  asymmetrically. Thus, the degree of centrality of interaction significantly affects the parameters of the spatial (angular) distribution of secondary particles \cite{4}-\cite{6}.

The initial state, about which it is usually very little direct experimental information, leads to significant peculiarities in the distribution of fragments and the multiplicity of secondary particles. The study of the peculiarities allows to give a physical interpretation of the results on the base of the differences in the initial states of collisions \cite{7}-\cite{11}. 

In the present work we analyze the peculiarities of the distribution of secondary particles depending on the degree centrality and the degree of asymmetry of the interacting nuclei on the base of experimental data of interactions of gold ions $^{197}Au$ at 10,7 AGeV with photoemulsion nuclei \cite{12}. 

The nuclear emulsion represents a certain combination of light and heavy nuclei. On the one hand, it allows to analyze various types of nuclear interactions obtained in exactly the same experimental conditions. On the other hand, it introduces additional problems with the identification of the target nucleus.

Peculiarities of distribution of secondary particles and fragments in depending on the initial conditions of collisions, was analyzed in two complementary directions. 

The first direction is the study of the parameters of the fragmentation of the projectile nucleus for separation of the peripheral interactions with a small number of interacting nucleons and with a large value of the total charge of the fragments of a projectile nucleus. 

The second direction was based on the study of the dependence of the number of fragments of the target nucleus and the multiplicity of secondary particles to assess the centrality degree of the interaction and to separate interactions with light and heavy emulsion nuclei.

\section{ Research of fragmentation parameters of projectile nucleus }

In interaction of huge gold with smaller nucleus of the photoemulsion the quantity of multi-charged fragments depends on centrality degree of the interaction. 

In Figure 1 the distributions of  total charge $Q$ of projectile nucleus fragments for events with various number of multi-charged fragments $N_f$ are presented.

\begin{figure}[h]
\begin{center}
\includegraphics*[width=0.6\textwidth,angle=0,clip]{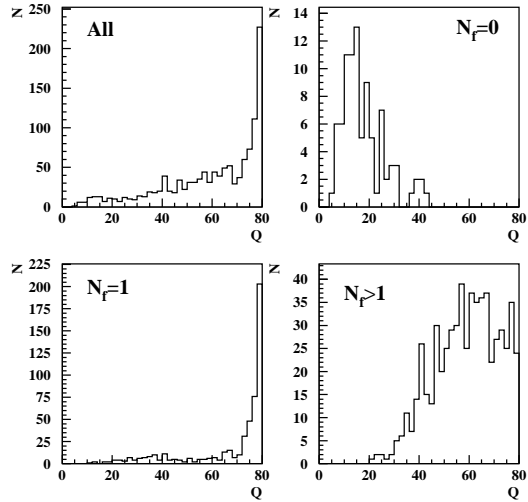}
\caption{\label{fig1} 
The distribution of the total charge $Q$ of the fragments of a projectile nucleus  for events with different numbers of multi-charged fragments $N_f$ for interactions of 10.7 AGeV $^{197}Au$ with emulsion nuclei }
\end{center}
\end{figure}

From Figure 1 it follows that in interactions with one multi-charged fragment, in the most of the events large total charge of the projectile nucleus fragments, is discovered.

The region of $Q>$70 is separated visually on a low background of other events. Thus, events with one multi-charged fragment and $Q>$70, probably, characterize the peripheral interaction. 

Events with several multi-charged fragments mainly are characterized by lower values of $Q$ and likely characterize a more central collisions. 

%Schematic visualization of events of varying degrees of centrality is presented% in Figure 2.
%   
%
%\begin{figure}[tbh]
%\begin{center}
%\includegraphics*[width=0.25\textwidth,angle=0,clip]{fig2a.eps}
%\includegraphics*[width=0.25\textwidth,angle=0,clip]{fig2b.eps}
%\includegraphics*[width=0.25\textwidth,angle=0,clip]{fig2c.eps}
%\caption{\label{fig2} 
%Schematic visualization of events of varying degrees of centrality
%}
%\end{center}
%\end{figure}

Special attention should be paid to the fact that there is a significant number of events in which  multi-charged fragments are absent. That is, under certain conditions,  the complete destruction of the huge gold at interaction with significantly less emulsion nuclei, is discovered. 

For the evaluation of multi-charged remnants the Figure 2 presents the distribution of total charge of multi-charged fragments of projectile nucleus $Q_f$ for events with different numbers of multi-charged fragments $N_f$.

\begin{figure}[tbh]
\begin{center}
\includegraphics*[width=0.6\textwidth,angle=0,clip]{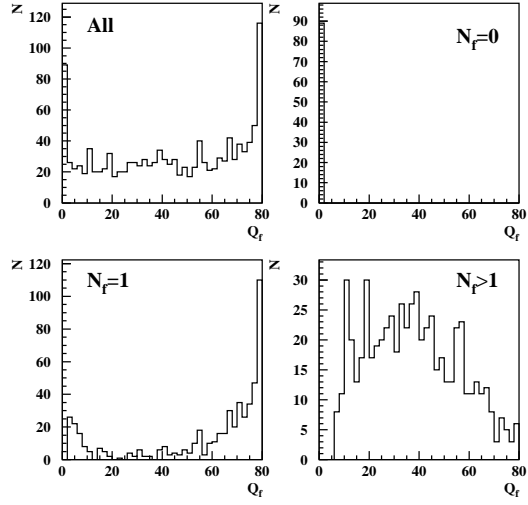}
\caption{\label{fig3} 
The distribution of the total charge of the multi-charged fragments of a projectile nucleus $Q_f$ for events with different numbers of multi-charged fragments $N_f$ for heavy-ion interactions of 10.7 AGeV $^{197}Au$ with emulsion nuclei
}
\end{center}
\end{figure}

As it can be seen from Figure 2 events with $N_f$ $>$ 1 fill the central part of the $Q_f$-distribution. In the event with $N_f$ = 1 in addition to the characteristic peak in the region of maximum values of $Q_f$, there is a significant number of events with low values of the residual nucleus.

 Research of the dependence of the number of fragments and the multiplicity of secondary particles

One of the most optimal parameters for assessing the centrality degree of interaction and separation of events with light and heavy emulsion nuclei, is the dependence of the number of fragments of the target nucleus and  the multiplicity of secondary particles ns. This correlation dependence is presented in figure 3.

\begin{figure}[tbh]
\begin{center}
\includegraphics*[width=0.6\textwidth,angle=0,clip]{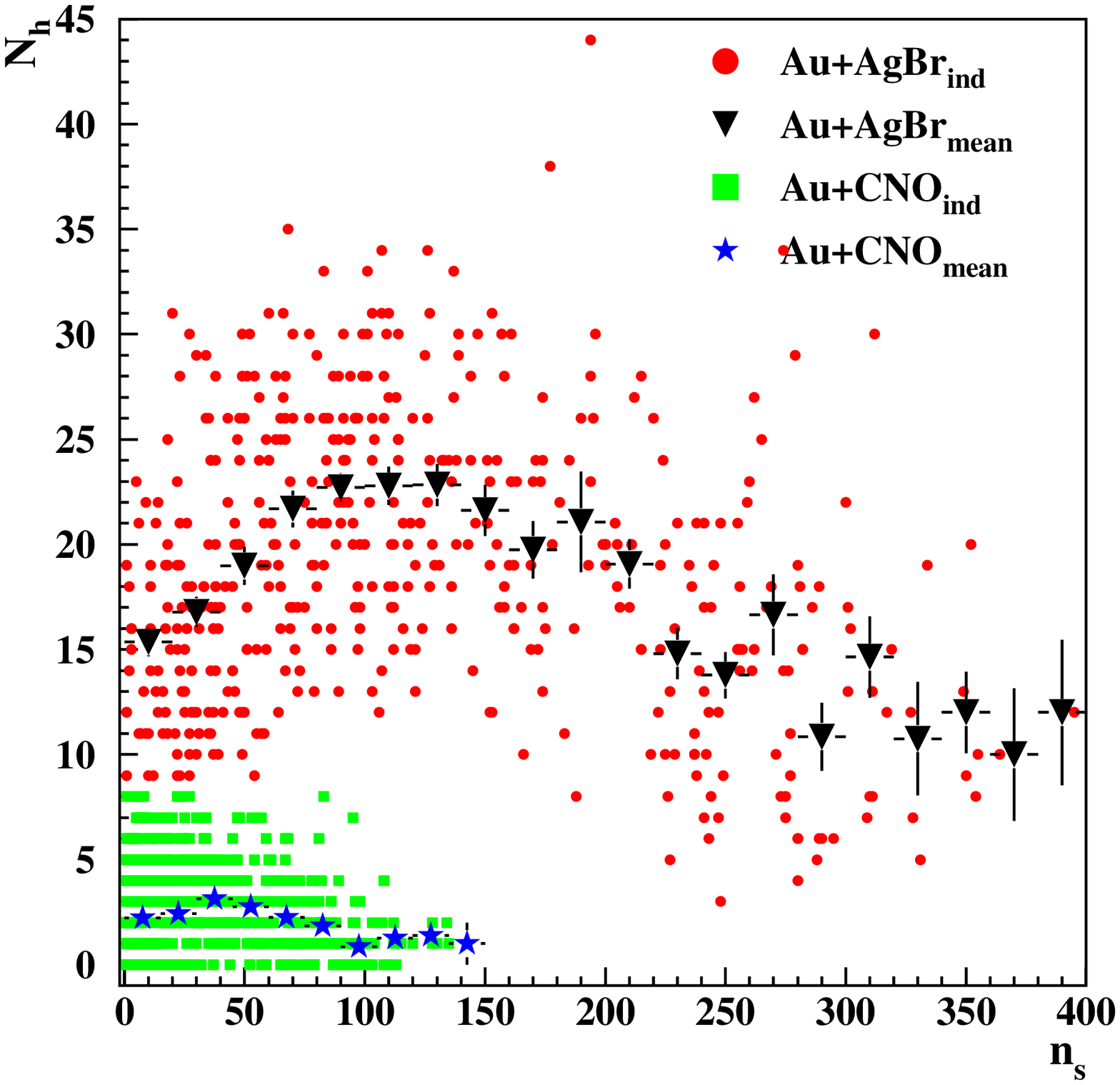}
\caption{\label{fig4} 
Correlation of the total number of fragments of the target nucleus $N_h$ and the  multiplicity  of secondary particles $n_s$ in interactions of 10.7 AGeV $^{197}Au$ with heavy ($AgBr$) and light ($CNO$) emulsion nuclei.
}
\end{center}
\end{figure}

As it can be seen from figure 3, events of interaction with light ($CNO$) and heavy ($AgBr$) emulsion nuclei can be separated quite well. Selection of interactions with light nuclei is limited to two requirements. 

First, the maximum number of fragments of the target nucleus may not exceed 8, which corresponds to the charge of the largest of light emulsion nuclei - nuclei of oxygen. 

Secondly, the maximum multiplicity ns in interactions with light nuclei emulsion is significantly lower compared to the interactions with heavy emulsion nuclei. Using of this fact allows to separate $Au+AgBr$ events with large multiplicity, in which the number of fragments of the target nucleus is less than 8, from $Au+CNO$ events.

For analysis of peripheral collisions in figure 3 the mean values of $N_h$ for groups of interactions with heavy and light emulsion nuclei, by described above criteria, are presented separately. 

In $Au+AgBr$ interactions the average dependence shows the steady growth in the field before $n_s$=110. Then it decreases and  at large $n_s>$250 it goes on the plateaus. Similar behavior shows and $Au+CNO$ dependence, but at a lower multiplicity: $n_s$=40 and $n_s>$100, respectively. This behavior reflects the degree of peripheral interaction. This assumption is confirmed by the difference in the total charge of the fragments of a projectile nucleus for the respective areas of $n_s$. This distribution of $Q$ is presented in figure 4. 

\begin{figure}[tbh]
\begin{center}
\includegraphics*[width=0.6\textwidth,angle=0,clip]{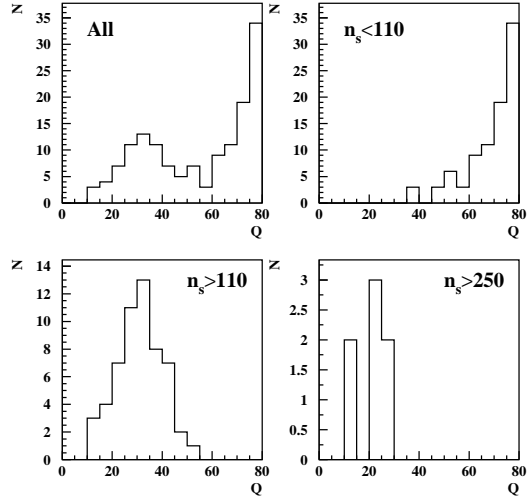}
\caption{\label{fig5} 
The distribution of the total charge of the projectile fragments, depending on the multiplicity ns for interaction $Au+AgBr$ with one multi-charged fragment $N_f$ =1.
}
\end{center}
\end{figure}

As it can be seen from figure 4 there is a clear two-peak distribution. The peak in the region of maximal values of $Q$ characterizes peripheral events with varying degrees of peripheral interaction on the rising branch of the average Nh-ns curve. The peak at low values of $Q$ characterizes the central event with varying degrees of centrality on the decreasing branch of the average $N_h$-$n_s$ curve.

\section {Conclusion}

In summary, we formulate the following main conclusions of the study of the peculiarities of the distribution of secondary particles depending on the degree centrality and the degree of asymmetry of the interacting nuclei.

The quantity of multi-charged fragments of projectile nucleus $N_f$ in interactions of sharply asymmetric nuclei essentially depends on the centrality degree. In the majority of peripheral interactions $N_f$  is equal 1. In the most of the central interactions there are $N_f >$ 1 or $N_f$ =0. 

$N_f$ =1 area is rather well divided in distribution of a total charge of projectile nucleus fragments $Q$. The peak in the field of the maximum values $Q$ characterizes peripheral events with various degree of the perifericos on a growing branch of average $N_h$-$n_s$ of a curve (characterizing dependence of number of fragments of the target nucleus and multiplicity of secondary particles from interaction area). And the peak in the field of small values $Q$ characterizes the central events with various degree of centrality on the decreasing branch of average $N_h$-$n_s$ of curve.

%\vspace{12pt}
%
%\begin{center}
%{\Large \bf Acknowledgements }
%\end{center}
%
%This work was supported by grant N1563/GF of Ministry of Education and Science %of Kazakhstan Republic.  

%+Bibliography

\end{document}